\pdfoutput=1
\documentclass[twocolumn,aps,prl,showpacs,superscriptaddress]{revtex4-1}
\usepackage[english]{babel} 
\usepackage[T1]{fontenc} 
\usepackage[latin1]{inputenc}
\usepackage[babel]{csquotes}
\usepackage{amsmath}
\usepackage{amsfonts} 
\usepackage{braket}
\usepackage{graphicx}
\usepackage{amsmath, amssymb, graphics, setspace}
\usepackage[colorlinks=true,citecolor=blue,urlcolor=blue]{hyperref}

\begin{document}

\title{Hybrid Entangled Entanglement in Vector Vortex Beams} 
\author{Vincenzo D'Ambrosio}
\affiliation{Dipartimento di Fisica, Sapienza
 Universit\`{a} di Roma, I-00185 Roma, Italy}
\author{Gonzalo Carvacho}
\affiliation{Dipartimento di Fisica, Sapienza
 Universit\`{a} di Roma, I-00185 Roma, Italy}
\author{Francesco Graffitti}
\affiliation{Dipartimento di Fisica, Sapienza
 Universit\`{a} di Roma, I-00185 Roma, Italy}
\author{Chiara Vitelli}
\affiliation{Dipartimento di Fisica, Sapienza
 Universit\`{a} di Roma, I-00185 Roma, Italy}
 \author{Bruno Piccirillo}
\affiliation{Dipartimento di Fisica, Universit\`{a} di Napoli Federico II, Complesso Universitario di Monte S. Angelo, 80126 Napoli, Italy}
\affiliation{CNR-SPIN, Complesso Universitario di Monte S. Angelo, 80126 Napoli, Italy}
\author{Lorenzo Marrucci}
\affiliation{Dipartimento di Fisica, Universit\`{a} di Napoli Federico II, Complesso Universitario di Monte S. Angelo, 80126 Napoli, Italy}
\affiliation{CNR-SPIN, Complesso Universitario di Monte S. Angelo, 80126 Napoli, Italy}
\author{Fabio Sciarrino}
\email{fabio.sciarrino@uniroma1.it}
\affiliation{Dipartimento di Fisica, Sapienza
 Universit\`{a} di Roma, I-00185 Roma, Italy}

\maketitle
 
{\bf Light beams having a vectorial field structure -- or polarization --  that varies over the transverse profile and a central optical singularity are called vector-vortex (VV) beams and may exhibit specific properties, such as focusing into ``light needles'' or rotation invariance, with applications ranging from microscopy and light trapping to communication and metrology. Individual photons in such beams exhibit a form of single-particle quantum entanglement between different degrees of freedom. On the other hand, the quantum states of two photons can be also entangled with each other. Here, we combine these two concepts and demonstrate the generation of quantum entanglement between two photons that are both in VV states -- a new form of quantum ``entangled entanglement''. This result may lead to quantum-enhanced applications of VV beams as well as to quantum-information protocols fully exploiting the vectorial features of light.}

Quantum entanglement \cite{Entanglement} lies at the basis of fundamental questions on the nature of reality, as exemplified by the Einstein-Podolsky-Rosen argument \cite{EPR} or the Schr\"{o}dinger's cat paradox \cite{gatto}. On the other hand entanglement is today also a key tool of quantum information technology, enabling applications such as quantum teleportation \cite{Teleportation}, superdense coding \cite{Coding} and quantum computation \cite{Computation}. Entangled quantum states are nowadays routinely produced with different physical systems spanning from atoms \cite{atom} to crystals \cite{crystal} and photons \cite{Photons}. Entangled photon pairs, in particular, are commonly generated by exploiting nonlinear optical processes \cite{Kwiat} and may show entanglement in several degrees of freedom such as frequency, path, orbital angular momentum (OAM), and polarization \cite{Pan}. 

The optical polarization -- defining the oscillation directions of the electromagnetic fields -- is typically approximately uniform in a light beam.
Yet, the polarization can also vary over the transverse profile, giving rise to vector beams with peculiar polarization patterns \cite{sing}.
The so-called vector vortex (VV) beams are a particular class of vector beams characterized by a central optical singularity surrounded by an azimuthally-varying pattern of polarization \cite{zhan2009,Car12}. These beams can be conveniently described as balanced nonseparable superpositions of polarization-OAM eigenmodes \cite{maurer2007tailoring}, with the OAM magnitude defining the ``order'' of the beams. Hence, photons in VV beams are actually entangled in these two degrees of freedom:  such kind of single-particle entanglement is also known as ``intrasystem'' entanglement and exists in classical systems too \cite{Aiello,forbes}.  VV beams have already found applications in areas ranging from microscopy \cite{abouraddy2006three} to metrology \cite{gear,Fate11}, optical trapping \cite{roxworthy2010optical}, nano-optics \cite{Neug14}, and quantum communication  \cite{DAmbrosio2012, vallone2014free, memory}. Due to their interesting properties, in the last years several techniques have been developed to generate, manipulate and analyze VV beams \cite{zhan2009,maurer2007tailoring,Car12,Fick14,damb15}.

In this paper, we report the generation and characterization of entangled pairs of VV photons of arbitrary order. In particular, we consider five combinations of VV mode orders, corresponding to different polarization patterns for the two beams. 
We simultaneously demonstrate, by a complete sixteen-dimensional quantum tomography, both the {\it intrasystem} entanglement between polarization and OAM within each photon and the {\it intersystem} entanglement between the two photon states: the former is related to the structure of VV states, the latter corresponds to entanglement between two complex vectorial fields that has, to our knowledge, never been reported before. In this way, we are demonstrating experimentally a novel form of ``entangled entanglement'' \cite{tee,eee}, namely \emph{ hybrid entangled entanglement} corresponding to two (intersystem) entangled pairs of (intrasystem) entangled pairs of qubits. 
Finally, by performing a non-locality test directly in the VV space, we show that entanglement between complex vectorial fields can be effectively exploited as a resource in fundamental quantum menchanics as well as quantum information.

\begin{figure}[htb]
\centering
\includegraphics[scale=0.31]{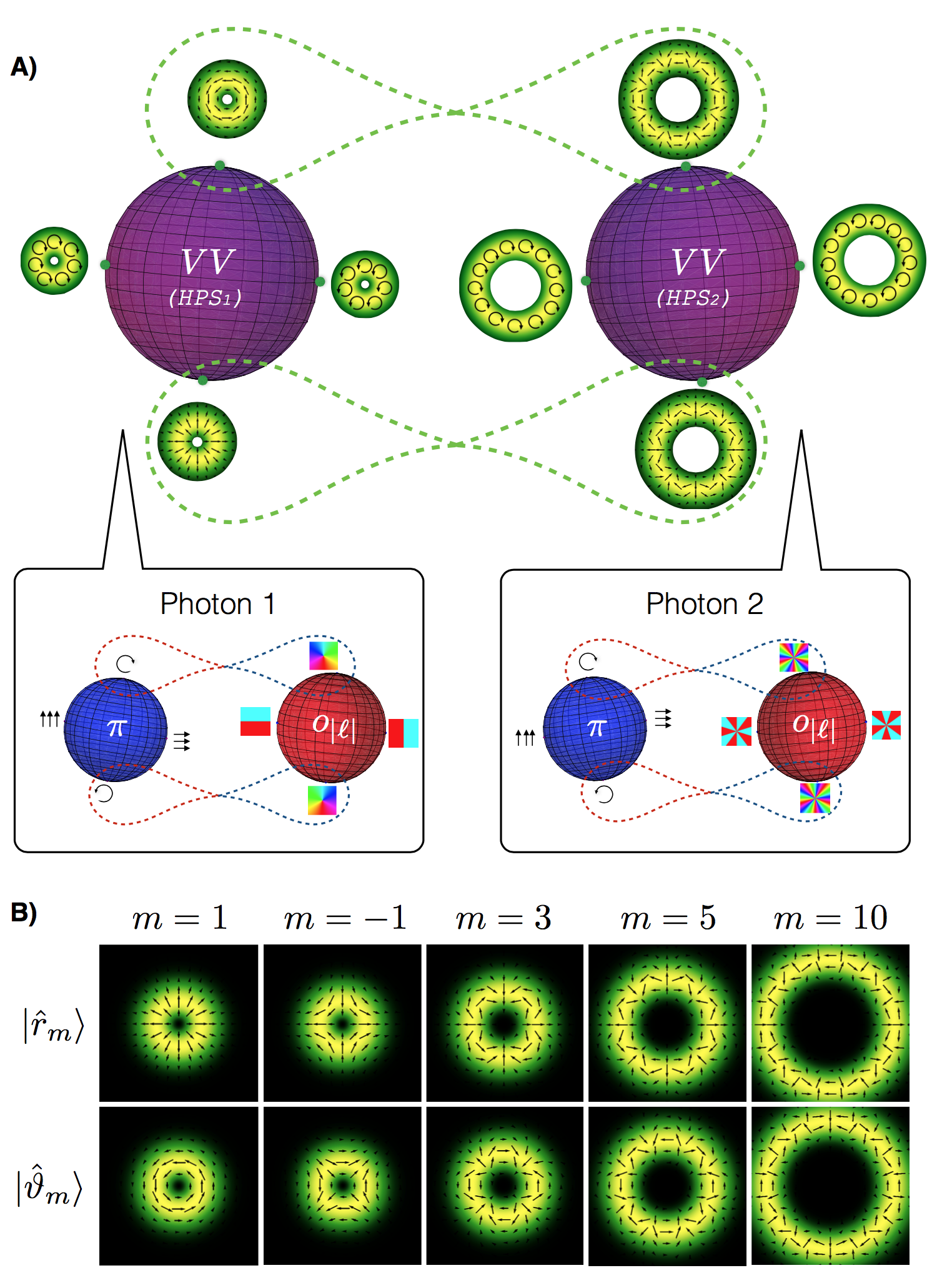}
\caption{Representation of entangled VV beams and their underlying hybrid entangled entanglement structure. Panel (A) shows the intersystem entanglement (green dashed lines) between VV beams belonging to different orders. Each VV state is defined on a hybrid-Poincar\'e sphere: purple spheres $HPS_1$ and $HPS_2$, corresponding to orders $m=1$ and $m=5$ respectively, represented together with the intensity and polarization patterns for the corresponding azimuthal, radial and circularly-polarized modes (for graphical reasons, these spheres are oriented with the two poles corresponding to polarization-OAM eigenstates aligned in the horizontal direction). The Poincar\'{e} spheres in the two lower boxes represent the polarization $\pi$ (blue spheres) and OAM $o_{|\ell |}$ (red spheres) spaces within each photon, with their respective polarization patterns and phase profiles. The intrasystem entanglement (blue/red dashed lines) between these two degrees of freedom of the single photons generates the VV states. Panel (B) reports the intensity and polarization patterns of radial and azimuthal VV beams of various orders.}
\label{modes}
\end{figure}

Let us denote with $\ket{R,\ell}$ ($\ket{L,\ell}$) the state of a photon with uniform right (left) circular polarization carrying $\ell\hbar$ of orbital angular momentum. A VV beam of order $m$ is defined in the two dimensional Hilbert space spanned by $\{\ket{R,m}, \ket{L,-m}\}$. In particular, we consider the two balanced superpositions $\ket{\hat{r}_m}=\frac{1}{\sqrt{2}}\left(\ket{R,m}+\ket{L,-m}\right)$, $\ket{\hat{\vartheta}_m}=\frac{1}{\sqrt{2}}\left(\ket{R,m}-\ket{L,-m}\right)$. For $m\neq0$, all these modes feature a polarization singularity and a null intensity in the center and hence exhibit the so-called doughnut beam profile. When analyzed through a linear polarizer, the corresponding intensity pattern consists of $2m$ petals \cite{Yao, gear}. When $m=1$, the states $\ket{\hat{r}_1}$ and $\ket{\hat{\vartheta}_1}$, corresponding to the well-known radially and azimuthally polarized beams \cite{zhan2009}, are also invariant under azimuthal rotations \cite{DAmbrosio2012,vallone2014free}. For brevity, we will refer to states $\ket{\hat{r}_m}$ and $\ket{\hat{\vartheta}_m}$ with the terms ``radial'' and ``azimuthal'', irrespective of $m$.  A generic VV beam (or photon VV state) can be represented on a ``hybrid Poincar\'{e} sphere'' (HPS) \cite{Holl11}, also known as ``higher-order Poincar\'{e} sphere'' \cite{milione2011higher}, where states $\{\ket{R,m},\ket{L,-m}\}$ lie on the poles and $\ket{\hat{r}_m}$ and $\ket{\hat{\vartheta}_m}$ lie on opposite points on the equator. Figure \ref{modes} shows examples of this representation and the polarization and intensity patterns of some of these modes. 
By noticing that the complete polarization-OAM Hilbert space of order $m$ is spanned by the four states $\{\ket{R,m}, \ket{L,-m}, \ket{L,m}, \ket{R,-m}\}$, we can also define $\pi$-modes of order $m$ as balanced superpositions in the Hilbert space spanned by $\{\ket{R,-m}, \ket{L,m}\}$:
 $\ket{\hat{\pi}^+_m}=\frac{1}{\sqrt{2}}\left(\ket{R,-m}+\ket{L,m}\right)$ and $\ket{\hat{\pi}^-_m}=\frac{1}{\sqrt{2}}\left(\ket{R,-m}-\ket{L,m}\right)$ 

In this work, we generate and detect VV modes by using a specialized optical component, the $q$-plate, a birefringent patterned slab that can couple/decouple polarization and OAM of single photons \cite{Marr06}. More in detail a $q$-plate with topological charge $q$ maps a photon with input state $\alpha\ket{R,0}+\beta\ket{L,0}$ into the output state
$\alpha\ket{L,-2q}+\beta\ket{R,2q}$ and vice versa. Thus, radial and azimuthal VV beams $\ket{\hat{r}_m}$ and $\ket{\hat{\vartheta}_m}$ are easily produced by using a linear horizontal (H) and vertical (V) input polarization, respectively, with $m=2q$. Superpositions of VV beams are also obtained from the corresponding input polarization superpositions. A $q$-plate can also be used to measure VV beams. Indeed, a radial (azimuthal) VV beam of order $m$ is transformed into a linear horizontal (vertical) uniformly polarized beam by a $q$-plate with $2q=m$. In this way, the measurement of a complex polarization pattern such as that of a VV beam is reduced to a much simpler polarization measurement, because the $q$-plate acts as an ``interface'' between the two dimensional space of VV beams of order $m$ and the polarization one \cite{Car12,damb15}. 
A more general and complete approach to analyze VV beams is by measuring separately the polarization and OAM, for instance by performing a quantum state tomography in the complete polarization-OAM Hilbert space.

\begin{figure}[htb]
\centering
\includegraphics[scale=0.2]{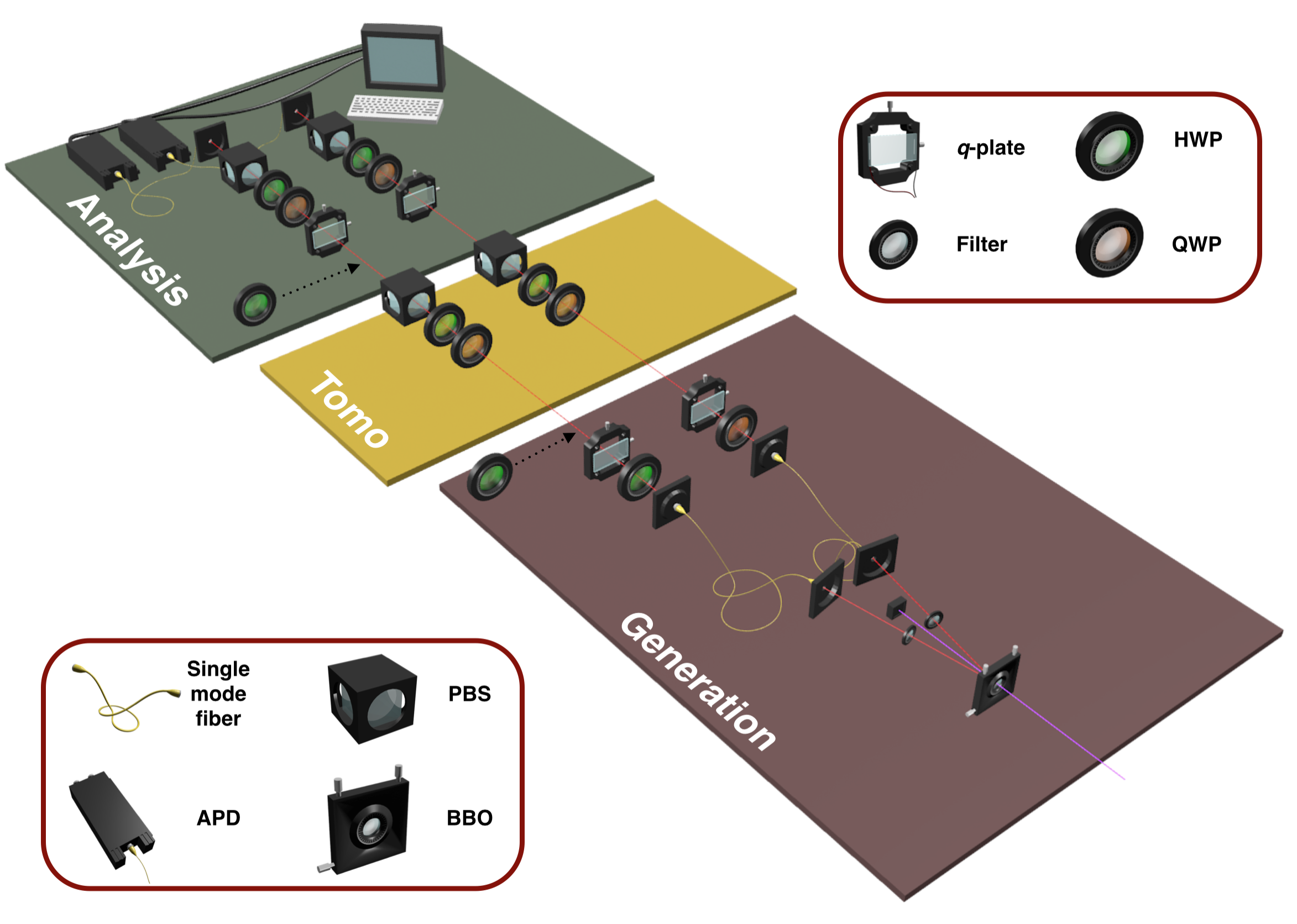}
\caption{Experimental apparatus for the generation and analysis of entangled VV photons. In the Generation stage, a polarization-entangled photon pair is produced in a BBO crystal and is then converted into a VV-entangled pair by two $q$-plates. The generation of a $\pi$-mode, needs an additional HWP after the $q$-plate. The Tomo section corresponds to a polarization-analysis stage formed by a quarter-wave plate (QWP), a HWP and a polarizer (PBS). This section is used only for the full tomography in the complete polarization-OAM space, while it is not needed for direct VV-space measurements. The Analysis stage deals with the OAM- or VV-mode analysis (it is OAM-mode if the Tomo-stage is inserted, otherwise it is VV-mode): on each arm a $q$-plate converts back the OAM- or VV-modes into uniform polarization states, which are then measured with a polarization analysis stage. The photons are then coupled into a single mode fiber that filters the non-Gaussian modes and are sent to single photon detectors.}
\label{setup}
\end{figure}

The experimental apparatus we used for the generation of entangled photons in VV modes is depicted in Fig. \ref{setup} (Generation section). A polarization-entangled photon pair is produced by exploiting spontaneous parametric down conversion in a $\beta$-barium borate crystal (BBO) \cite{Kwiat}. The generated state is 
$\frac{1}{\sqrt{2}}\left(\ket{H}_1\ket{V}_2-\ket{V}_1\ket{H}_2\right)$, where the subscript denotes the photon output path. After the crystal, both photons are filtered in wavelength, via a narrowband interference filter, and in spatial mode, via a single-mode fiber. This last operation sets the transverse mode of both photons to a fundamental Gaussian one ($TEM_{00}$), corresponding to vanishing OAM. As a last step, two q-plates with topological charges $q_1$ and $q_2$, transform the two polarization photon states into VV modes of orders $m_1=2q_1$ and $m_2=2q_2$, respectively.

The resulting state emerging from the source is the following:
\begin{equation}
\ket{\Psi^-_{m_1,m_2}}=\frac{1}{\sqrt{2}}\left(\ket{\hat{r}_{m_1}}_1\ket{\hat{\vartheta}_{m_2}}_2-\ket{\hat{\vartheta}_{m_1}}_1\ket{\hat{r}_{m_2}}_2\right),
\end{equation}
which corresponds to a pair of entangled vector-vortex beams of orders $m_1$ and $m_2$. A complete set of VV Bell states can be obtained by performing local operations on one of the photons. Indeed an input $\ket{\Phi^\pm}$ state in polarization is transformed by the two $q$-plates into $\ket{\Phi^\pm_{m_1,m_2}}$ and $\ket{\Psi^+}$ is transformed into $\ket{\Psi^+_{m_1,m_2}}$. The same applies for any linear combination of Bell states: this versatile technique allows to generate arbitrary maximally entangled vector vortex beams.

To prove the generality of our approach, we generated various combinations of VV mode orders. By exploiting $q$-plates with $q=1/2, 3/2, 5/2$ and $5$ we considered the following combinations ($m_1,m_2$): (1,1), (1,5) (1,10) and (3,5). Finally, we generated the pair (1,-1). The VV modes with $m=-1$ are $\pi$-modes \cite{Roxw10}, which can be obtained by flipping the circular polarization handedness with a half-wave plate (HWP) added after the $q$-plate with $q=1/2$, or by directly exploiting a $q$-plate with $q=-1/2$. 

To fully analyze the entangled pairs, we performed a quantum state tomography in the polarization and the OAM subspaces corresponding to VV mode of each photon. In this way it is possible to measure both the intrasystem and the intersystem entanglement of our states, hence we can certify the generation of hybrid entangled entanglement. 
The corresponding experimental setup is shown in Fig.\ \ref{setup} (Tomo and Analysis sections). After a polarization analysis stage (two waveplates and a polarizer), each photon is sent to a $q$-plate and a second polarization analysis stage. In this configuration, the $q$-plate transfers the information initially written in the OAM subspace into the polarization state of the photon that can be then analyzed with standard techniques \cite{nagali2009quantum,transferrer}. An hyper-complete set of measurements of polarization and OAM for both photons  (overall 1296 settings) has been performed to fully reconstruct the density matrix of the entangled photon pair in the sixteen-dimensional Hilbert space (2 dimensions for polarization and 2 for OAM for each photon). Such characterization has been carried out for all the entangled pairs of different VV modes. The comparison between experimental and theoretical density matrices is reported in Fig. \ref{tomo}A for two VV  combinations. The complete set of experimental density matrices can be found in the Supplementary Material.
The quality of our states can be measured by calculating the fidelity $F=Tr[\sqrt{\sqrt{\rho_{theo}}\rho_{exp}\sqrt{\rho_{theo}}}]$ between the experimental density matrix $(\rho_{exp})$ and the corresponding theoretical one ($\rho_{theo}$). The average fidelity is: $F=0.97\pm 0.01$.

In order to quantify the intrasystem entanglement of VV states, we calculated the concurrence $C$ for the single photon reduced density matrix in the polarization-OAM space after any projective measurement performed on the other photon. Figure \ref{tomo}B shows the concurrence distribution for each photon when the other is projected over 34 different states $\{\chi_i\}$. As expected, the distribution is divided in two regions (blue and orange) corresponding to entangled ($C=1$) and separable ($C=0$) states respectively. Indeed when $\chi_i$ contains a circular polarized state or a OAM eigenstate the state of the other photon is separable while in all the other cases it is a fully entangled state corresponding to a vector-vortex field.

\begin{figure*}[htb]
\centering
\includegraphics[scale=0.28]{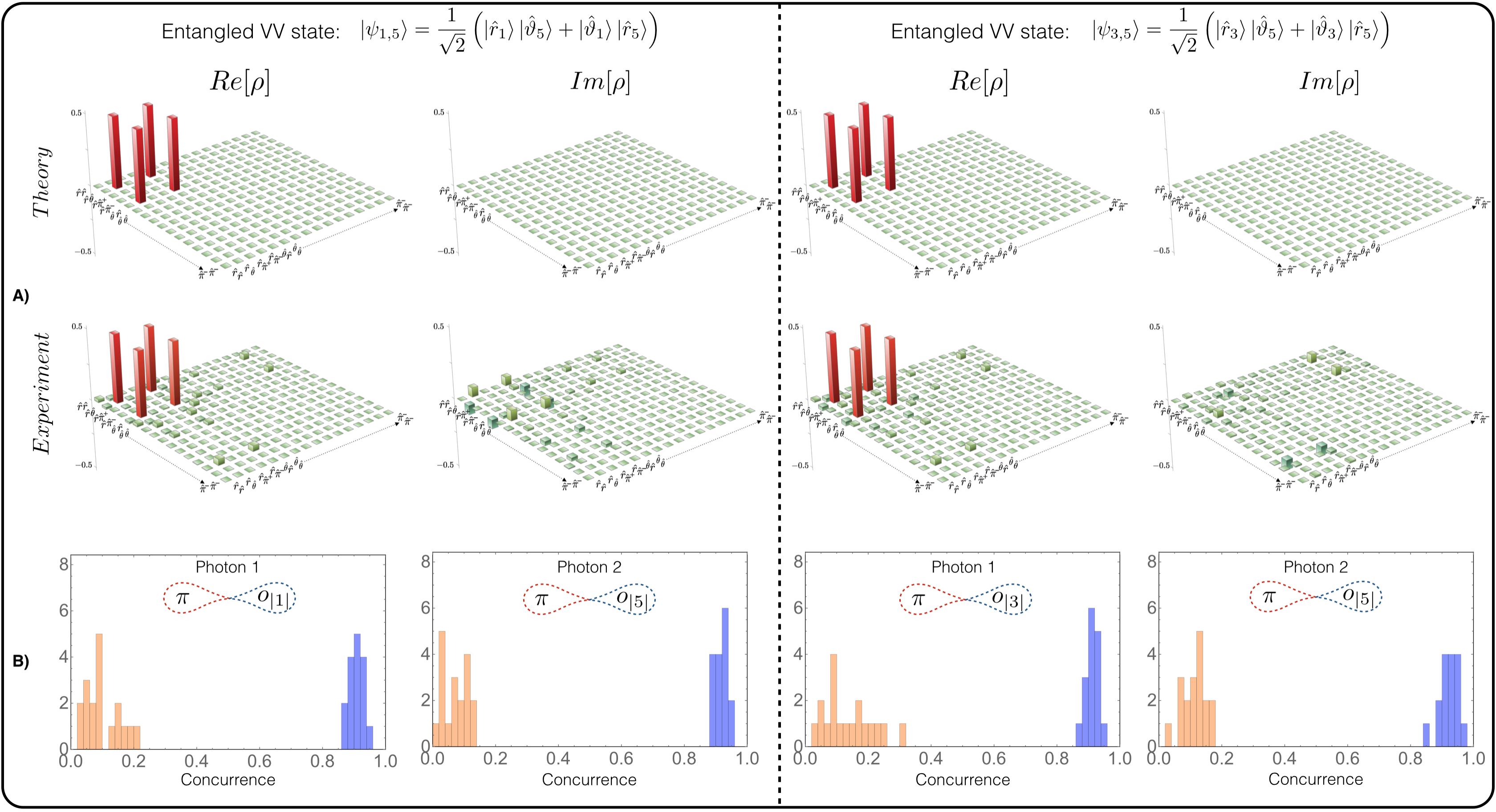}
\caption{Experimental results. A) Experimental and theoretical density matrices for entangled VV beams of orders $(m_1=1,m_2=5)$ (on the left) and $(m_1=3,m_2=5)$ (on the right). B) Polarization-OAM concurrence distributions. To quantify the intrasystem entanglement we calculated the concurrence distributions for the single photon reduced density matrix after 34 different projective measurements performed on the other photon. All the distributions are divided in two regions  corresponding to entangled (blue) and separable (orange) states.}
\label{tomo}
\end{figure*}

On the other hand, the intersystem entanglement between two VV beams can be quantified by directly calculating the concurrence $C$ for the corresponding density matrix in the VV space.
The average value for the concurrence over the five VV couples considered here is $C= 0.91 \pm 0.03$ (see also Supplementary Materials). 

Beside the intriguing fundamental aspects of the entanglement, this feature is also a keystone in quantum based technologies. 
One of the most common ways to certify the entanglement is through a violation of a Bell inequality. Hence we performed a non-locality test for each pair of VV photons in order to show that entanglement between complex vectorial fields constitutes an exploitable resource in quantum protocols. 

More in detail, we can map the bidimensional VV space of order $m$ into a qubit encoding by defining $\ket{0}=\ket{\hat{r}_m}$ and $\ket{1}=\ket{\hat{\vartheta}_m}$. To test the entanglement we then measured the correlations between the two photons in the VV space. In a standard quantum information scenario, two parties (Alice and Bob) perform measurements on one photon on the entangled pair respectively. According to this convention we performed a violation of the CHSH inequality \cite{CHSH} $S=|E(a_0,b_0)+E(a_1,b_0)+E(a_0,b_1)-E(a_1,B_1)|\leq 2$ where $a_i$($b_j$) are the outcomes of Alice's (Bob's) measurement settings $i$($j$) and $E(a_i,b_j)$ are the correlators between Alice and Bob measurements that quantify the probability of observing a coincidence in Alice and Bob detectors for a given measurement configuration.
The maximum violation of the CHSH inequality allowed by quantum theory is given by $S=2\sqrt{2}$ corresponding to a maximally entangled state and to the following measurement bases: $\{\ket{0},\ket{1}\}$ and $\frac{1}{\sqrt{2}}\{\ket{0}+\ket{1},\ket{0}-\ket{1}\}$ for Alice and $\{\cos(\pi/8)\ket{0}+\sin(\pi/8)\ket{1},-\sin(\pi/8)\ket{0}+\cos(\pi/8)\ket{1}\}$ and  $\{\sin(\pi/8)\ket{0}+\cos(\pi/8)\ket{1},-\cos(\pi/8)\ket{0}+\sin(\pi/8)\ket{1}\}$ for Bob. Being defined in the VV space, these measurement settings need the ability to perform projective measurements directly in the VV qubit Hilbert space. To this purpose, as mentioned above, we exploited a $q$-plate as an interface between VV and polarization spaces (Analysis stage in Fig. \ref{setup}). For a given VV order $m$, a q-plate with $q=m/2$ followed by a HWP and a polarizer (PBS) allows one to perform a projective measurement on a state of the form $\cos(\gamma)\ket{0}+\sin(\gamma)\ket{1}$, for any value of $\gamma$. Different measurement settings for Alice and Bob correspond to different angles of their HWPs. 
The experimental values of $S$ are reported in Table \ref{CHSH} and correspond to a violation of CHSH inequality by 39 to 73 standard deviations, without any correction due to dark counts. 

\begin{table}
\begin{tabular}{|c|c|c|}
\hline\hline
{\bf $(m_1,m_2)$} &  S (raw data) & {\bf $S_{corr}$}\\
\hline
(1,1)& 2.654 $\pm$ 0.009 & 2.727 $\pm$ 0.009  \\
(1,5)& 2.649 $\pm$ 0.013 & 2.738 $\pm$ 0.014  \\
(1,10)& 2.437 $\pm$ 0.010 & 2.591 $\pm$ 0.011  \\
(3,5) & 2.621 $\pm$ 0.016 & 2.716 $\pm$ 0.017  \\
(1,-1) & 2.592 $\pm$ 0.010 & 2.664 $\pm$ 0.010\\
\hline\hline
\end{tabular}
\caption{Measured S parameter for CHSH inequalities using different orders of VV beams (first column) for raw data and for data corrected for dark counts (second and third column respectively). }\label{CHSH}
\end{table}

In this work we have fully investigated the hybrid entangled entanglement in a photonic system composed of two VV beams. Such a system shows indeed two different types of entanglement: an  intrasystem entanglement between polarization and OAM of each photon is responsible for the complex polarization pattern of VV beams, the two photons are also entangled with each other via an  intersystem entanglement which lies at the basis of the non-locality concept. We investigated the structure of our system by performing a full state tomography and quantified both types of entanglement by calculating the concurrence for the reduced density matrices in the single photon and VV modes Hilbert spaces respectively. Finally we have performed a non-locality test to prove that intersystem entanglement between entangled subsystem can be used as a resource in quantum protocols despite the complexity of the two subsystems involved. This study can pave the way towards a quantum enhancement in VV beams applications as well as the realization of quantum-information protocols that can take advantage of the combined action of both intrasystem and intersystem entanglement. 

We thank G. Rubino for her contribution in the preliminary stage of the experiment.
This work was supported by PRIN (Programmi di ricerca di rilevante interesse nazionale) project AQUASIM and ERC-Starting Grant 3D-QUEST (3D-Quantum Integrated Optical Simulation; Grant Agreement No. 307783): www.3dquest.eu.

\end{document}